\documentclass[a4paper,11pt]{article}

\usepackage{jinstpub} 

\usepackage{lineno}

\usepackage{graphicx}
\usepackage{subfigure}
\usepackage{color,xcolor}
\usepackage{stfloats}
\usepackage{hyperref}

\title{\boldmath Simulation study of particle identification using cluster counting technique for the \BESthree\, drift chamber}

\author[a,b,1]{Shuiting Xin\note{Corresponding author.},}
\author[a,b,1]{Guang Zhao,}
\author[a,b]{Linghui Wu,}
\author[a,b]{Mingyi Dong,}
\author[a,b]{Gang Li,}
\author[a,b]{Shengsen Sun,}
\author[a,b,c]{Xinchou Lou}
\affiliation[a]{Institute of High Energy Physics, 19B, Yuquan Road, Shijing District, Beijing 100049, China}
\affiliation[b]{University of Chinese Academy of Sciences (CAS), 19A Yuquan Road, Shijing District, Beijing 100049, China}
\affiliation[c]{Physics Department, University of Texas at Dallas, Richardson TX, USA.}

\emailAdd{xinshuiting@ihep.ac.cn, zhaog@ihep.ac.cn}

\abstract{The particle identification of charged hadrons, especially for the separation of $K$ and $\pi$, is crucial for the flavour physics study. Ionization measurement with the cluster counting technique, which has much less fluctuation than traditional $dE/dx$ measurement, is expected to provide better particle identification for the \BESthree\, experiment. Simulation studies, including a Garfield++ based waveform analysis and a performance study of $K/\pi$ identification in the \BESthree\, offline software system have been performed. 
The results show that $K/\pi$ separation power and PID efficiency would be improved significantly in the momentum range above 1.2 GeV/c using cluster counting technique even with conservative resolution assumption.
}


\begin{document}
\maketitle
\flushbottom

\section{Introduction}\label{sec:1}

The Beijing Spectrometer (\BESthree)~\cite{20097} detector at the Beijing Electron Positron Collider (BEPC\uppercase\expandafter{\romannumeral2}) is a general purpose detector designed for studies of hadron physics and $\tau$-charm physics. It has collected data around 40 fb$^{-1}$ at different center-of-mass energies from 2 to 5 GeV since 2009.  
To consider further physics opportunities and extend the physics potential of BES\uppercase\expandafter{\romannumeral3}, an upgrade of BEPC\uppercase\expandafter{\romannumeral2} was initiated~\cite{BESIII:2020nme}. For this upgrade, high luminosity and increasing beam energy were proposed and discussed. Hence, excellent performance of particle identification (PID) is essential for both high precision measurement and new physics searches for the future programme. 

The current PID of \BESthree\, is based on the momentum and ionization energy loss ($dE/dx$) measurements by the multilayer drift chamber (MDC) and the time of flight measurement by time-of-flight (TOF) counters. The $dE/dx$ resolution is around 6\%. The time resolution of TOF is 68 ps in the barrel region and 60 ps in the end cap region~\cite{Shengsen_2012}~\cite{Guo2017}. 
Given the need for a high accurate charged particle identification, the cluster counting ($dN/dx$) technique~\cite{4329616}, an alternative method of ionization measurement, can be a choice for the \BESthree\, drift chamber.

Cluster counting has been demonstrated in experiments for several decades~\cite{4329616}. When a charged particle passes through a drift chamber, a sequence of clusters of one or more ion-electron pairs emerge along the track. The Landau distribution with an infinite long tail describes the fluctuations of energy loss. On the other hand, the primary ionization itself allows a preferable understanding of the ionization behavior of gas, because the Poisson nature of the number of clusters provides a smaller uncertainty than a Landau distribution. Therefore, counting the primary clusters can conceivably reach an intrinsic resolution of the ionization process, and offer an improved particle separation capability. However, cluster counting has not been achieved in past high energy physics experiments due to limitations on electronics. In recent years, the development of electronics technology facilitates realization of the cluster counting method. It has been proposed in the IDEA drift chamber~\cite{Abada2019} and is also being studied for CEPC drift chamber~\cite{cepc}.

In this paper, we performed a Monte Carlo (MC) study of applying the cluster counting technique for PID in the \BESthree\, drift chamber. To estimate the $dN/dx$ resolution, a Garfield++ based simulation is carried out. A further study of $K/\pi$ identification capability is practiced in the \BESthree\, offline software system (BOSS). This paper proceeds as follows: Section~\ref{sec:2} introduces the simulation study using Garfield++ software~\cite{Garfield}. Section~\ref{sec:3} presents the Monte Carlo study in BOSS~\cite{boss}. Section~\ref{sec:4} gives a brief conclusion.

\section{Garfield based simulation}\label{sec:2}
A two-step simulation work is performed in this study.
In order to probe the potential of cluster counting technique, a theoretical estimation of ionization measurement is firstly obtained from Garfield++ simulation on ionization process. The study of $dE/dx$ and $dN/dx$ behaviors is illustrated in section~\ref{subsec:1}. For the second step, the resolution degradation of $dN/dx$ is considered in the simulation of waveform processing. The waveform analysis in described section~\ref{subsec:2}. 

\subsection{Ionization simulation for $dE/dx$ and $dN/dx$}\label{subsec:1}

The ionization simulation is to model the energy loss and ionized clusters of charged tracks under the condition of MDC. The simulation of $dE/dx$ and $dN/dx$ for MDC are framed with Garfiled++ program and interfaced to Heed. The program Heed~\cite{Pfeiffer:2018yam}, an implementation of the photon-absorption ionization (PAI) model simulates the ionization produced along the track of charged particles. 

\begin{figure}[h]
\centering
  \subfigure[$dE/dx$]{\includegraphics[width=.40\linewidth]{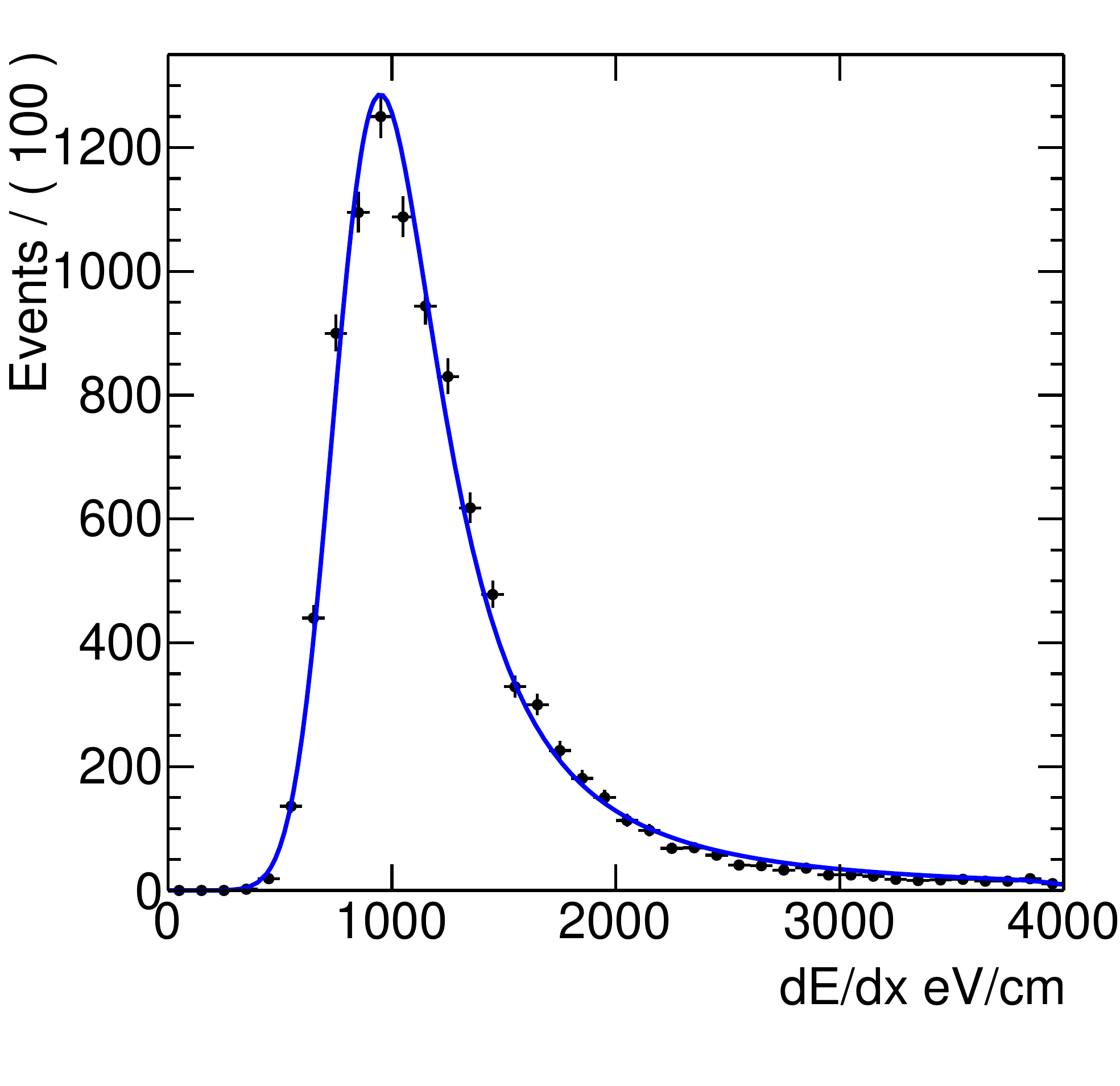}\label{fig:sub-cell:a}}
  \subfigure[$dN/dx$]{\includegraphics[width=.40\linewidth]{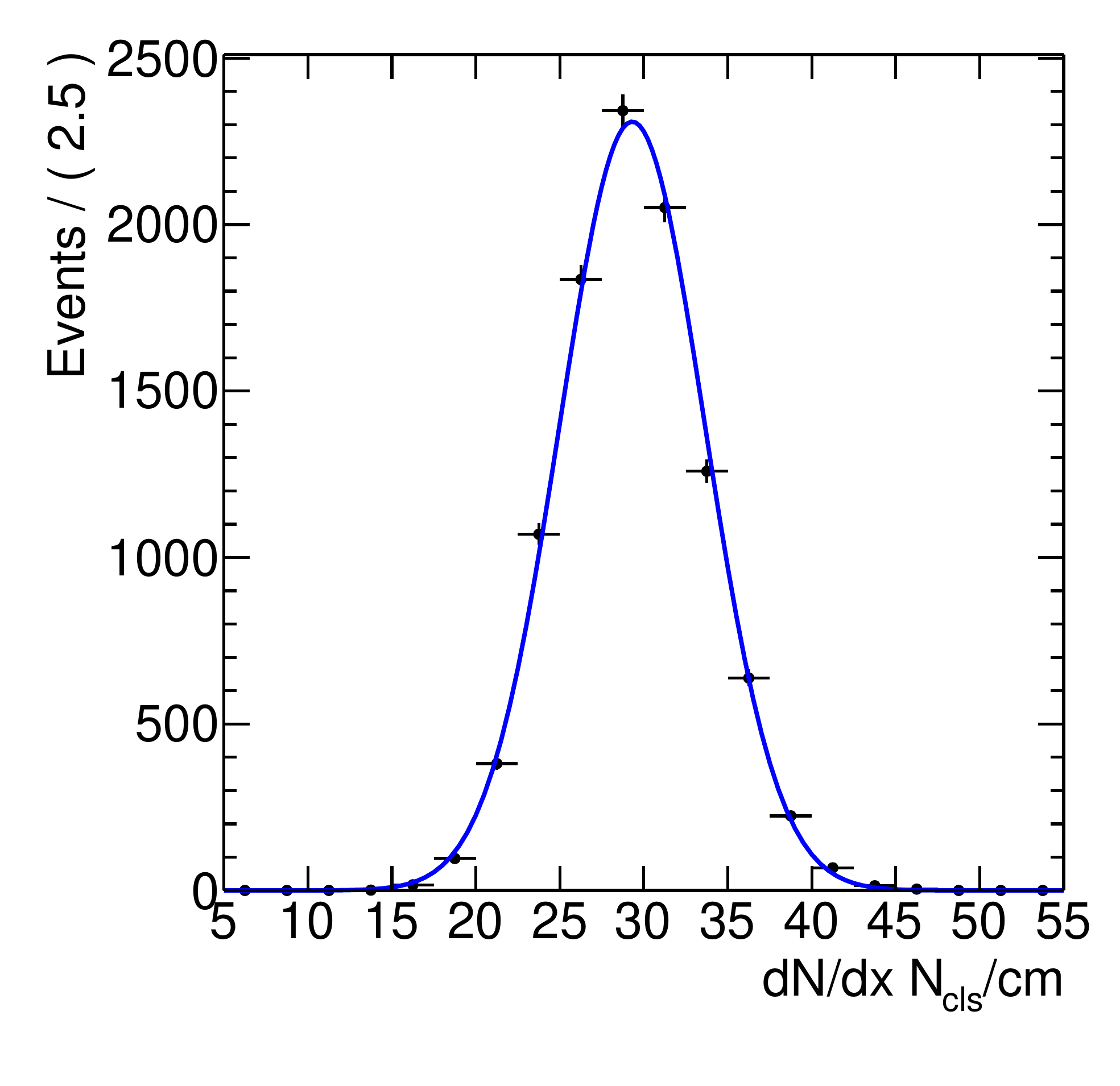}   \label{fig:sub-cell:b}}
  \caption{$dE/dx$ and $dN/dx$ distribution of 2 GeV/c $\pi$ for one cell. }
  \label{fig:sub-cell}
\end{figure}

In the simulation, the configuration of detector geometry and material of MDC are implemented in Garfield++ program.  The drift chamber consists of 43 cylindrical layers, the cell size is around 12mm $\times$ 12 mm for the inner eight layers and 16.2 mm $\times$ 16.2 mm for the outer layers. The chamber is filled with a helium based mixture He-C$_3$H$_8$ 60:40  and operates in a 1T magnetic field~\cite{20097}. Figure~\ref{fig:sub-cell} shows the $dN/dx$ and $dE/dx$ distribution of 2 GeV/c $\pi$ and $K$ of one cell, separately. As expected, the energy loss distribution is peaked with a long tail, as shown in Figure~\ref{fig:sub-cell:a}, where a fit is performed using a Landau convoluting a Gauss function. The $\delta$ electrons received enough energy from the incident particle ionize other atoms and result in such events in the tail of energy loss distribution. While the $dN/dx$ distribution is a symmetric Gaussian, independent of that extra ionization process. 

\begin{figure}[h]

  \centering
  \subfigure[$dE/dx$]{\includegraphics[width=.40\linewidth]{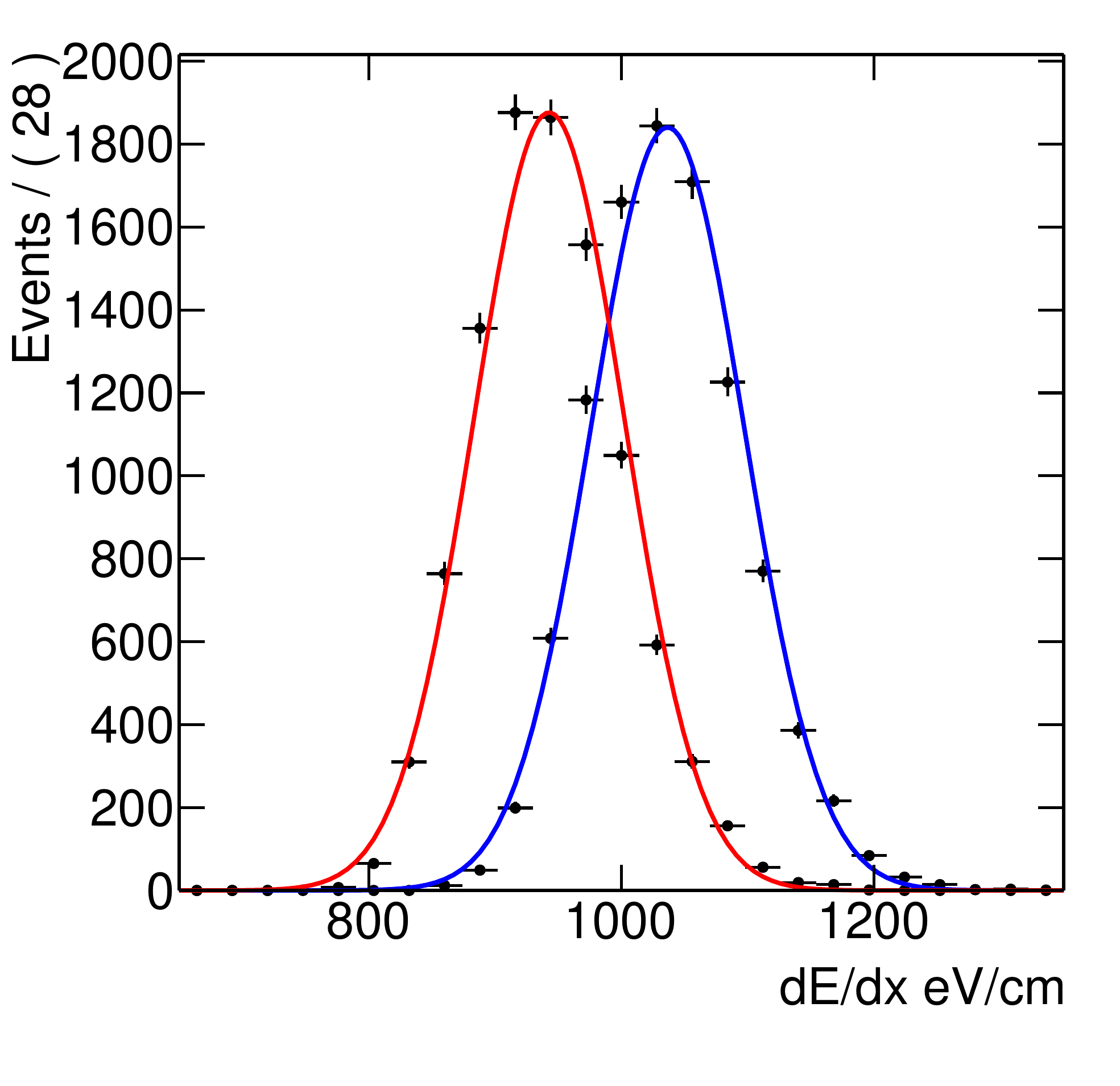}\label{fig:sub-track:a}}
  \subfigure[$dN/dx$]{\includegraphics[width=.40\linewidth]{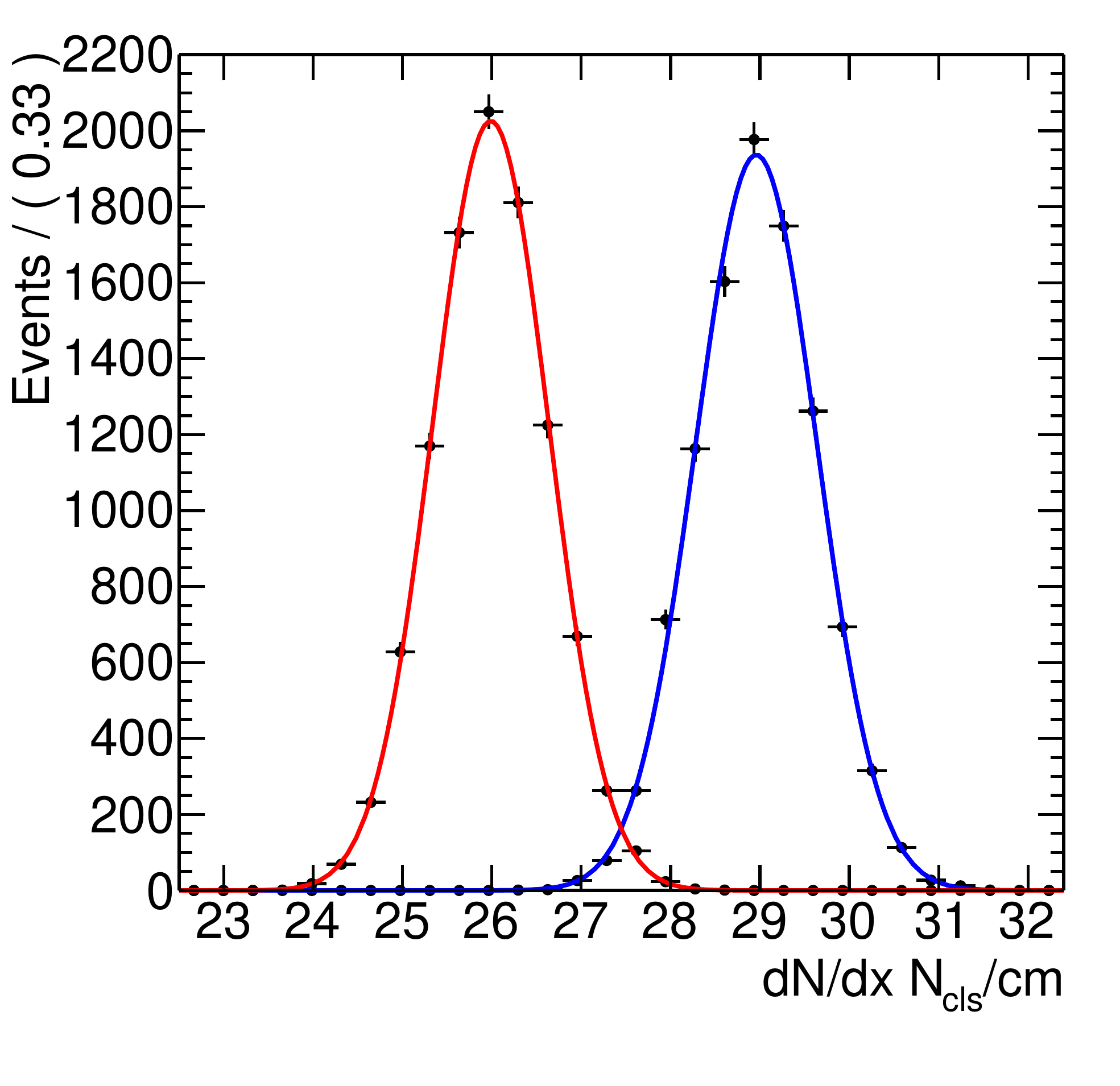}\label{fig:sub-track:b}}
  \caption{$dE/dx$ and $dN/dx$ distribution of 2 GeV/c $\pi$ (Left) and $K$ (Right) for one track. }
  \label{fig:sub-track}
\end{figure}

For a track that passes through MDC, $dE/dx$ is obtained using the truncated mean method over all hits to reduce the impact of the long Landau tail, as displayed in Figure~\ref{fig:sub-track:a}. $dN/dx$ is averaged over all hits in the track, as shown in Figure~\ref{fig:sub-track:b}. The $dE/dx$ resolution defined as $\frac{\sigma_{dE/dx}}{dE/dx}$ is about 5.6\% at this momentum. While $dN/dx$ resolution defined as $\frac{\sigma_{dN/dx}}{dN/dx}$, reaches a value below 3\%, which is a factor of 2 better than $dE/dx$.

\begin{figure}[h]
    \centering  
    \subfigure[$dE/dx$ of $\pi$ and $K$]{\includegraphics[width=0.49\linewidth]{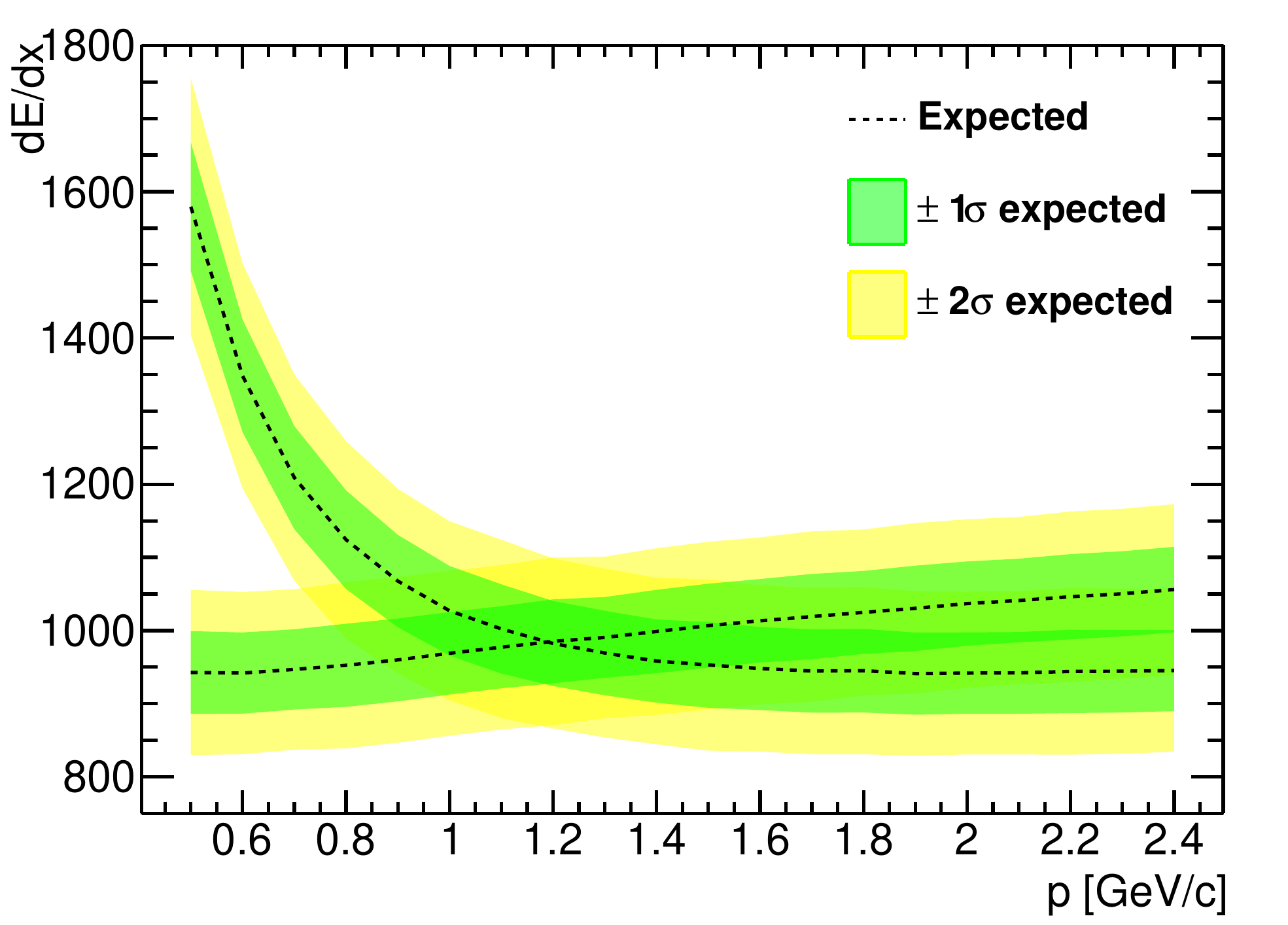}}
    \subfigure[$dN/dx$ of $\pi$ and $K$]{\includegraphics[width=0.49\linewidth]{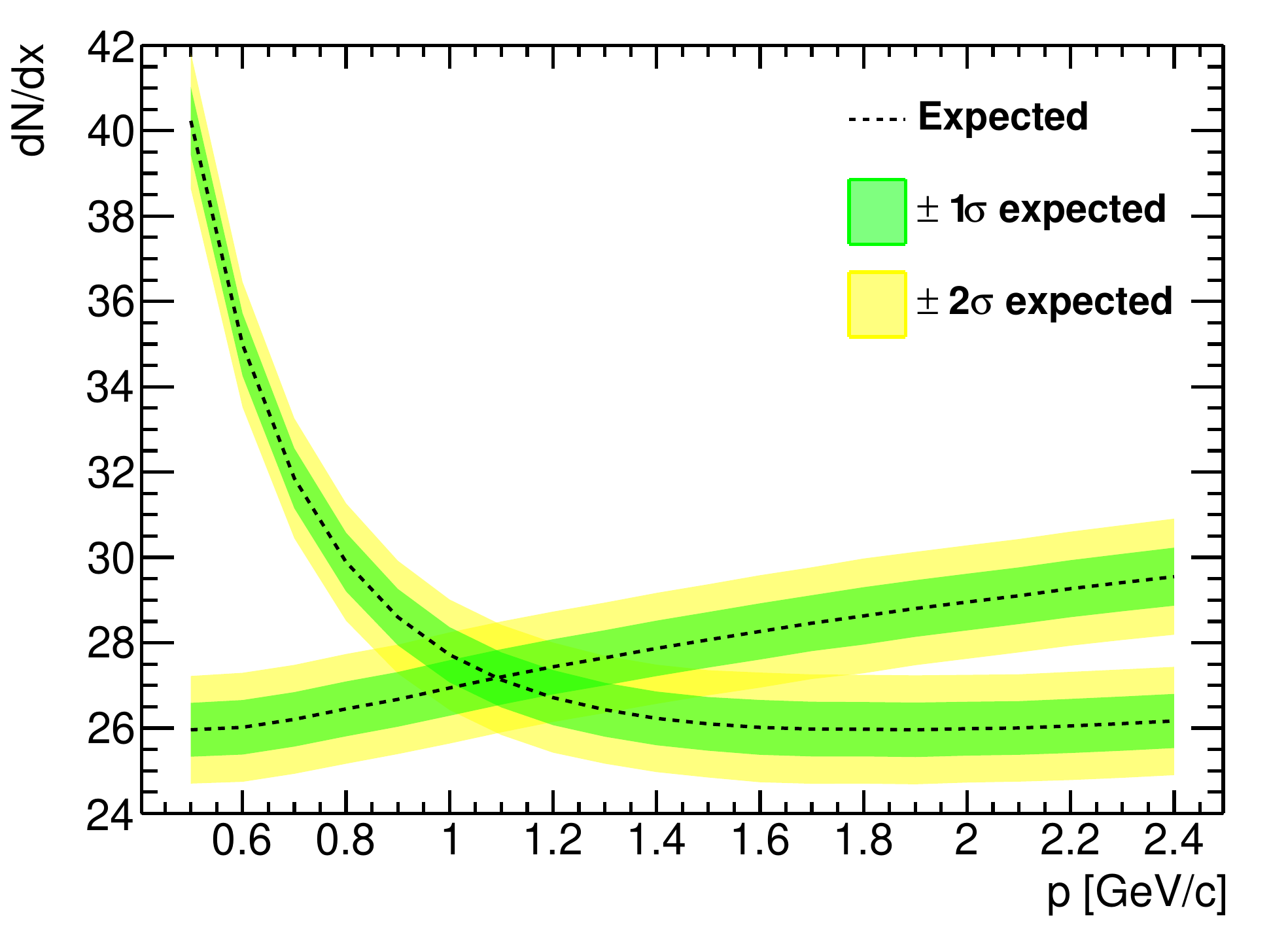}}
    \caption{$dE/dx$ and $dN/dx$ versus momentum of $\pi$ and $K$, where the $\pi$ is falling from top to bottom. The error bands are drawn with 1$\sigma$ and 2$\sigma$ uncertainty from a Gaussian fit.}
    \label{fig:4}
\end{figure}

Figure~\ref{fig:4} shows the evolution of those two quantities along the momentum of $\pi$ and $K$. $dN/dx$ apparently has narrow bands of $K$ and $\pi$ than $dE/dx$, giving a clear trend on separation. Apart from that, the intersection point of $\pi$ and $K$ around 1.1 GeV/c, infers that two particle species are indistinguishable at that momentum. As a consequence, neither energy loss measurement nor cluster counting is ineffectual at all in the region overlapped within 1 $\sigma$ band. In this case, the identification of the two particle species has to be achieved with the TOF. 

\subsection{$dN/dx$ simulation based on waveform processing}\label{subsec:2}

The critical content of cluster counting technique implementation is the waveform processing. It composes the waveform digitization of signal produced by individual ionization, waveform analysis with suitable algorithm applied to detect peaks associated to clusters. 

\begin{figure}[!ht]
    \centering
    \includegraphics[width=0.99\linewidth]{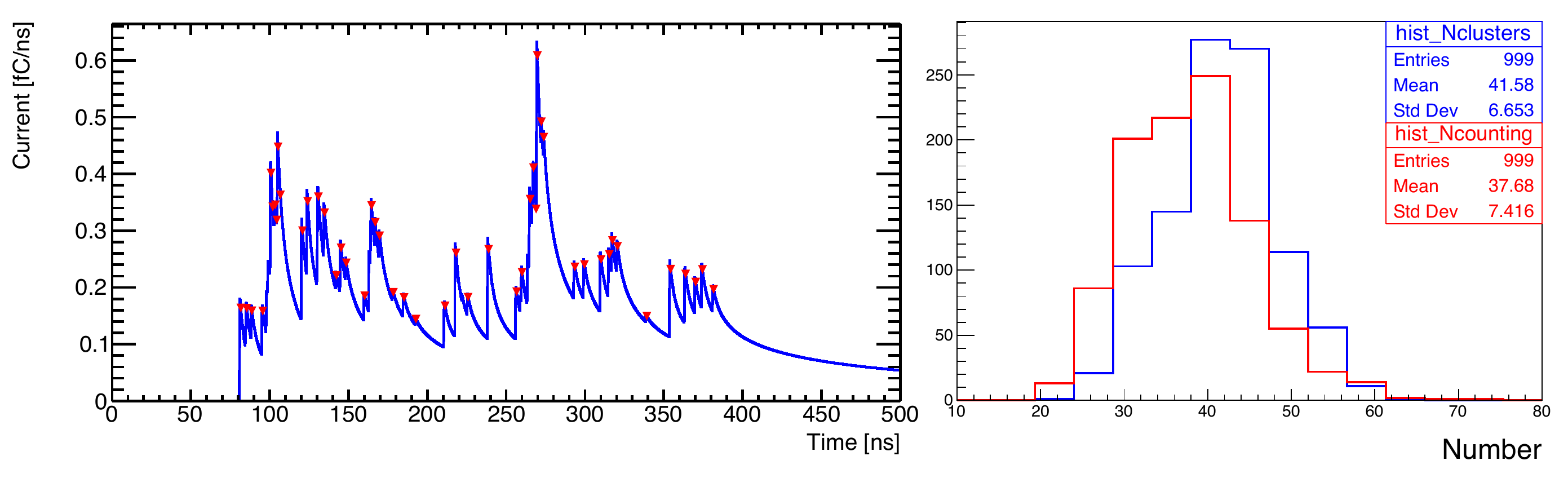}
    \caption{Left: A typical waveform of induced signal, a time window 500 ns is set, the red triangles are the peaks found by TSpectum; Right: Comparision of the distribution of original primary cluster and found peaks:  primary ionization (blue)  and found peaks (red) for simulated events in one cell.}
    \label{fig:3}
\end{figure}

As described in Section~\ref{subsec:1}, the drift and avalanche of electron-ion pairs are simulated for event in one cell. Collecting all the charges by the electrodes, the induced current is shaped as a sequence of failing pulses. A typical induced signal manufactured by Garfield++ program is shown in Figure~\ref{fig:3}. 
A peak finding algorithm named TSpectrum, provided by ROOT~\cite{root}, is adapted for identification of peaks. The histograms in Figure~\ref{fig:3} show the number of truth clusters and the candidate clusters from algorithm for a statistic of one thousand events, respectively. Compared with the truth distribution, the mean value of distribution from cluster counting decreases by 10\%, and the sigma is degraded from 16\% to 19.6\% for a single cell. No pronounced fake peaks are observed in the peak finding task. The main reason for the inefficiency of counting is high cluster density. For the gas mixture of He-C$_3$H$_8$ 60:40, the cluster density of the primary ionization is 26 per cm for Minimum Ionizing Particle (MIP) and 29 per cm for 2 GeV/c $\pi$. As the pileup region shown in Figure~\ref{fig:3}, some clusters are very close in time on the waveform due to cluster density and longitudinal diffusion, which makes peak finding difficult, and leads to a worse resolution.

There are some more factors that cause the degradation of the resolution in the experiment: 1) The sampling rate of the electronics 
has a substantial impact on the digitization process so as on the peak finding efficiency. 2) Noise makes  miscounting from the actual peaks. 3) Secondary ionization causes overestimation of the primary counts. In order to estimate degradation considering all those effects, 30\% and 60\% degradation of $dN/dx$ resolution are proposed.

A PID performance study is carried out with the $dN/dx$ considering the resolution degradation. To quantify the PID performance, the separation power $S^M_{K,\pi}$ between $K$ and $\pi$ of measurement $M$ is introduced, where $M$ refers to 
the substitutions of $dE/dx$, $dN/dx$ quantity.
\begin{equation}
    S^M_{K,\pi} = \frac{|\overline{M_{K}}-\overline{M_{\pi}}|}{\sqrt{\sigma(M_{K})^2+\sigma(M_{\pi})^2}}. 
\end{equation}

In the equation, $\overline{M_{K}}$ and $\overline{M_{\pi}}$ indicate the average value of the distribution of measurement $M$, $\sigma(M_{K})$ and $\sigma(M_{\pi})$ are the corresponding standard deviation of the measurement $M$ for $K$ and $\pi$, respectively. 

\begin{figure}[h]
    \centering
    \includegraphics[width=0.80\linewidth]{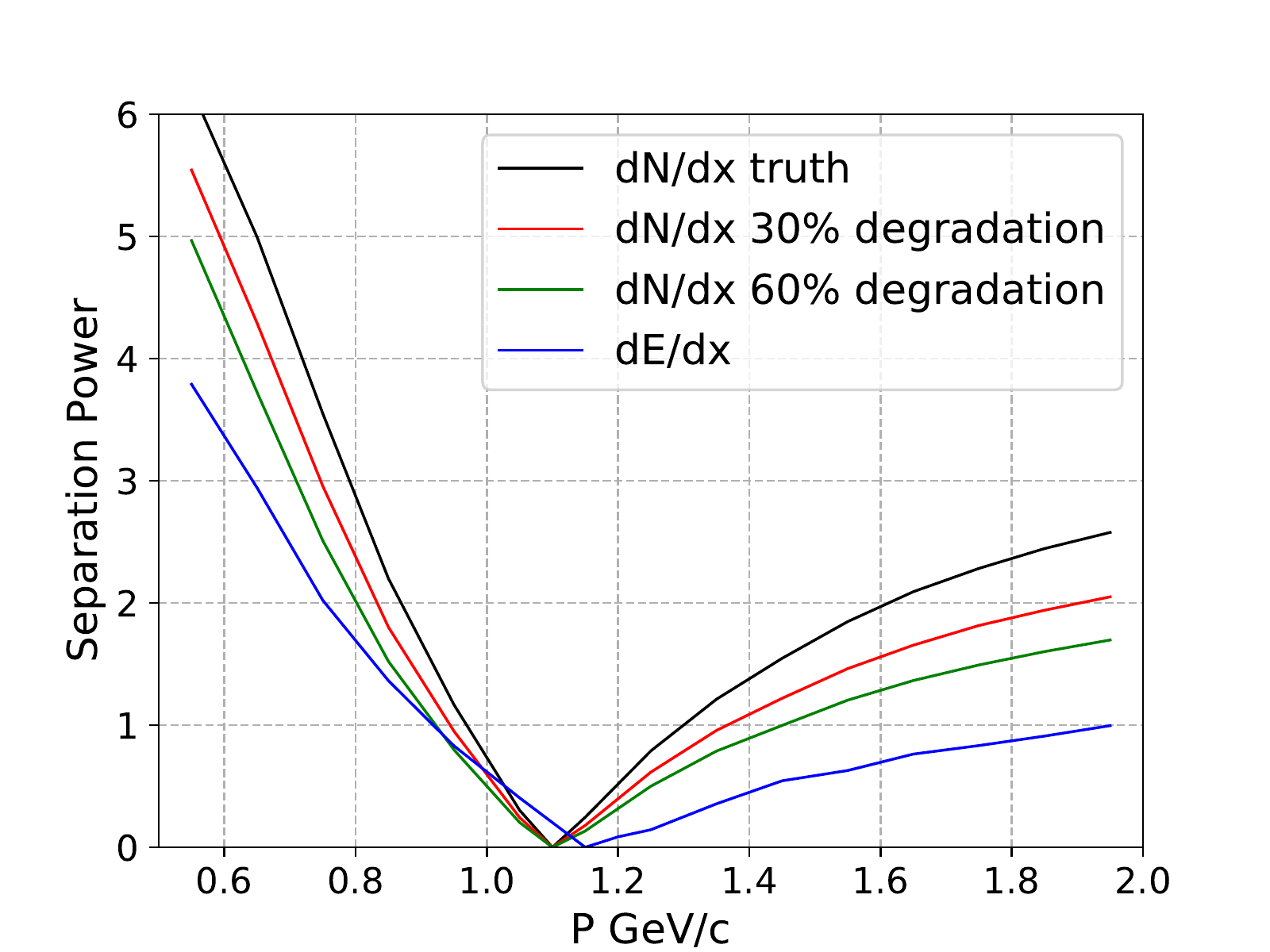}
    \caption{Separation power of $\pi/K$ as a function of momentum for  truth $dN/dx$, 30\% degradation of $\sigma_{dN/dx}$, 30\% degradation of $\sigma_{dN/dx}$ and truth $dE/dx$.}
    \label{fig:7}
\end{figure}

The separation power as a function of momentum is drawn in Figure~\ref{fig:7} for $dE/dx$ and  $dN/dx$. The $K/\pi$ separation with cluster counting surpasses energy loss measurement for almost whole range of momentum, except for momenta of particle from 0.9 GeV/c to 1.2 GeV/c, which could be recovered by the TOF detector. It is found that the resolution of $dN/dx$ is the crucial point in determining adequate PID performance. A gain of 250\% is seen from the separation. In the most degradation scenario, $dN/dx$ still delivers an advantage over $dE/dx$ in the high momenta region, which offers an increase of about 170\%. 

\section{Monte-Carlo study in the BES\uppercase\expandafter{\romannumeral3 } offline software system}\label{sec:3} 

In this section, a MC study is performed in the \BESthree\, offline software system (BOSS). The BOSS utilizes full offline data processing including detector simulation, reconstruction and calibration of sub-detectors and analysis toolkit. A $dN/dx$ model based on parameterization method is implemented in the analysis toolkit of BOSS. The $dN/dx$ estimation is modeled from the previous simulation at track level. The predicted cluster counts are expressed as a function of $\beta\gamma$ of the charged track. $dN/dx$ is assigned to event for particle assumption, $\pi, K$, according to their $\beta\gamma$ for the next step of probability calculation. 

In order to apply the PID analysis, probability variables are calculated for each particle hypothesis. To involve the contribution of TOF, time of flight measurement is considered in $M$ from now on. The procedure for the calculation of particle probability is following:

\begin{enumerate}
    \item For each particle hypothesis, we calculate a statistic variable $\chi^{M}_{i}$ for measurement $M$, with expected mean value and deviation:  $\chi^{M}_i = \frac{M_{i} - M_{i,exp}}{\sigma_{M_{i}}} (i = \pi, K)$
    \item Find the sum of square of  $\chi^M$ : $\chi^2_{i} = \sum (\chi^M_{i})^2  $
    \item Derive the probability of each particle hypothesis for a certain $\chi^2_i$: $P_i(\chi^2_i|n)$, which donates the upper tail probability of Chi-squared distribution of degree $n$.
\end{enumerate}

The particle type is determined by the hypothesis with the highest probability. For instance, a charged track is identified as $K$ if $P_{K} > P_{\pi}$, and vice versa. Consequently, the PID efficiency of a particle $x$ ($K$ or $\pi$) can be defined,
\begin{align}
  \epsilon_{x} = \frac{N_{x \rightarrow x}}{N_{x}},  
\end{align}
where $N_{x}$ is the total number of generated $x$, $N_{x \rightarrow x}$ is the number of correctly identified $x$.

Single particle events of $K$ and $\pi$ are simulated for PID efficiency study. The Monte Carlo samples are produced with GEANT4\cite{AGOSTINELLI2003250}, which include the geometry description of the BES\uppercase\expandafter{\romannumeral3} detector and the detector response. Only events having exactly single track candidate with good quality are selected.   

\begin{figure*}[ht]
\centering
\subfigure[$K$ efficiency without TOF]{\includegraphics[width=0.48\linewidth]{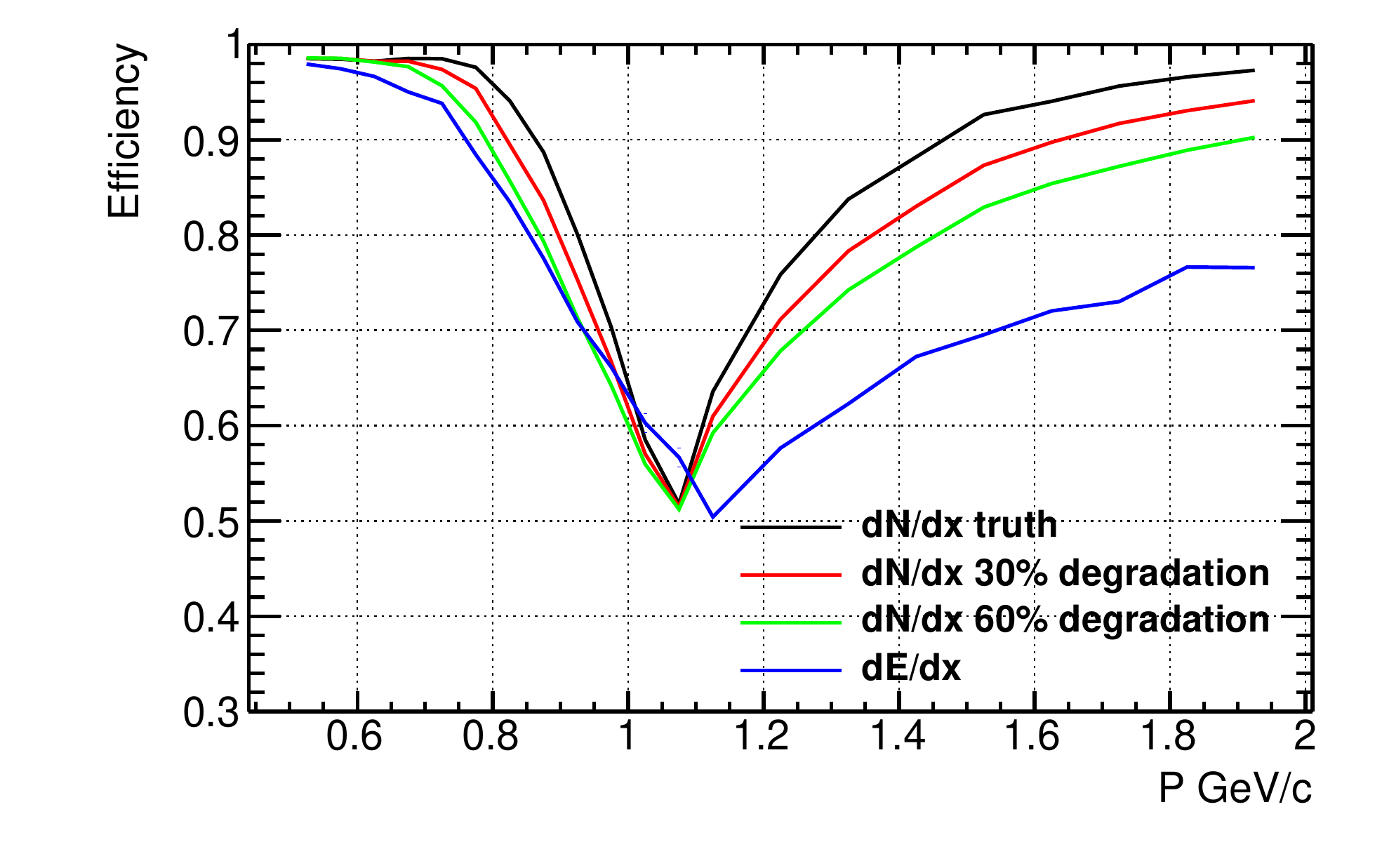}\label{fig:6:1}}
\subfigure[$K$ efficiency with TOF]{\includegraphics[width=0.48\linewidth]{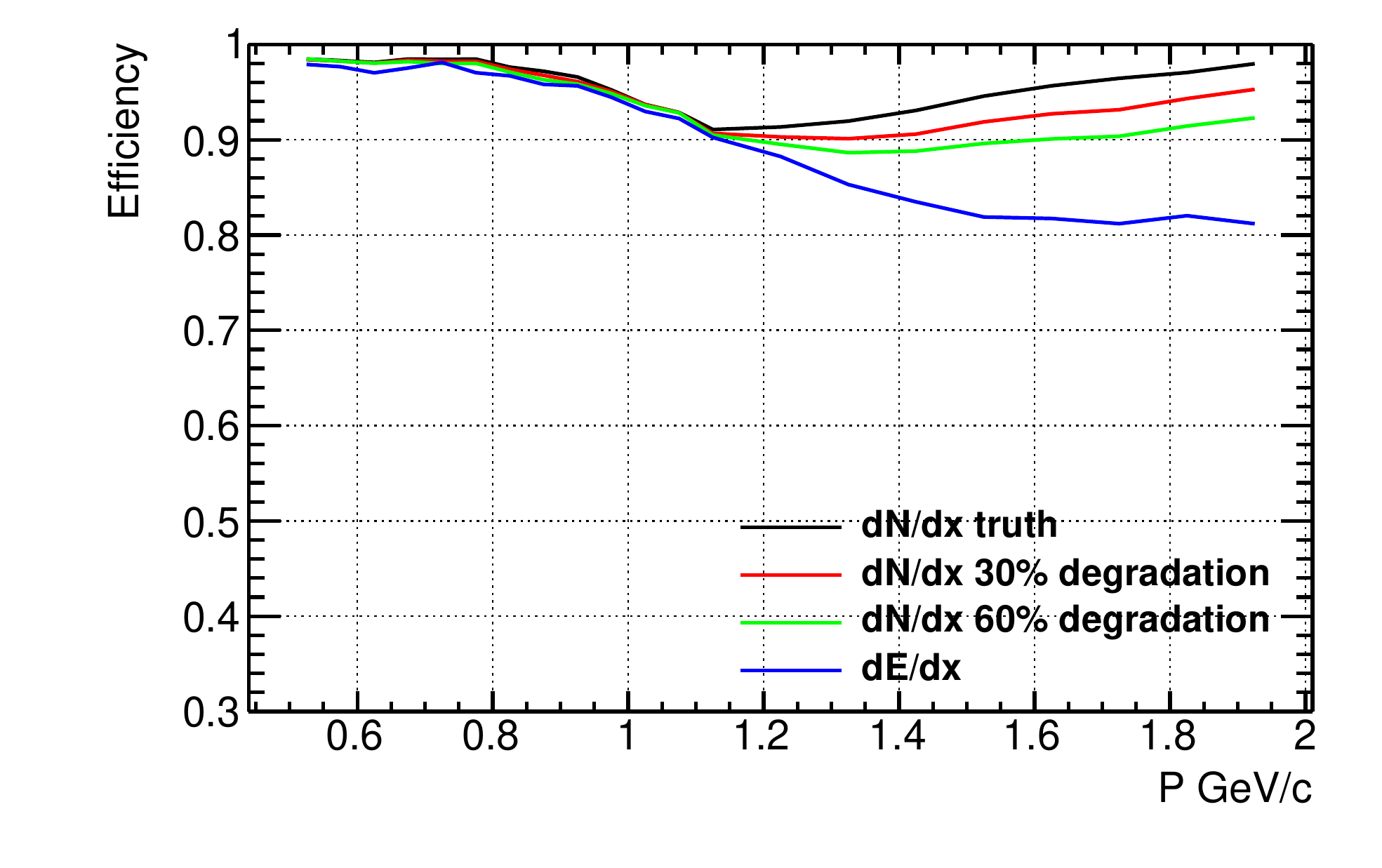}\label{fig:6:2}}
\subfigure[$\pi$ efficiency without TOF]{\includegraphics[width=0.48\linewidth]{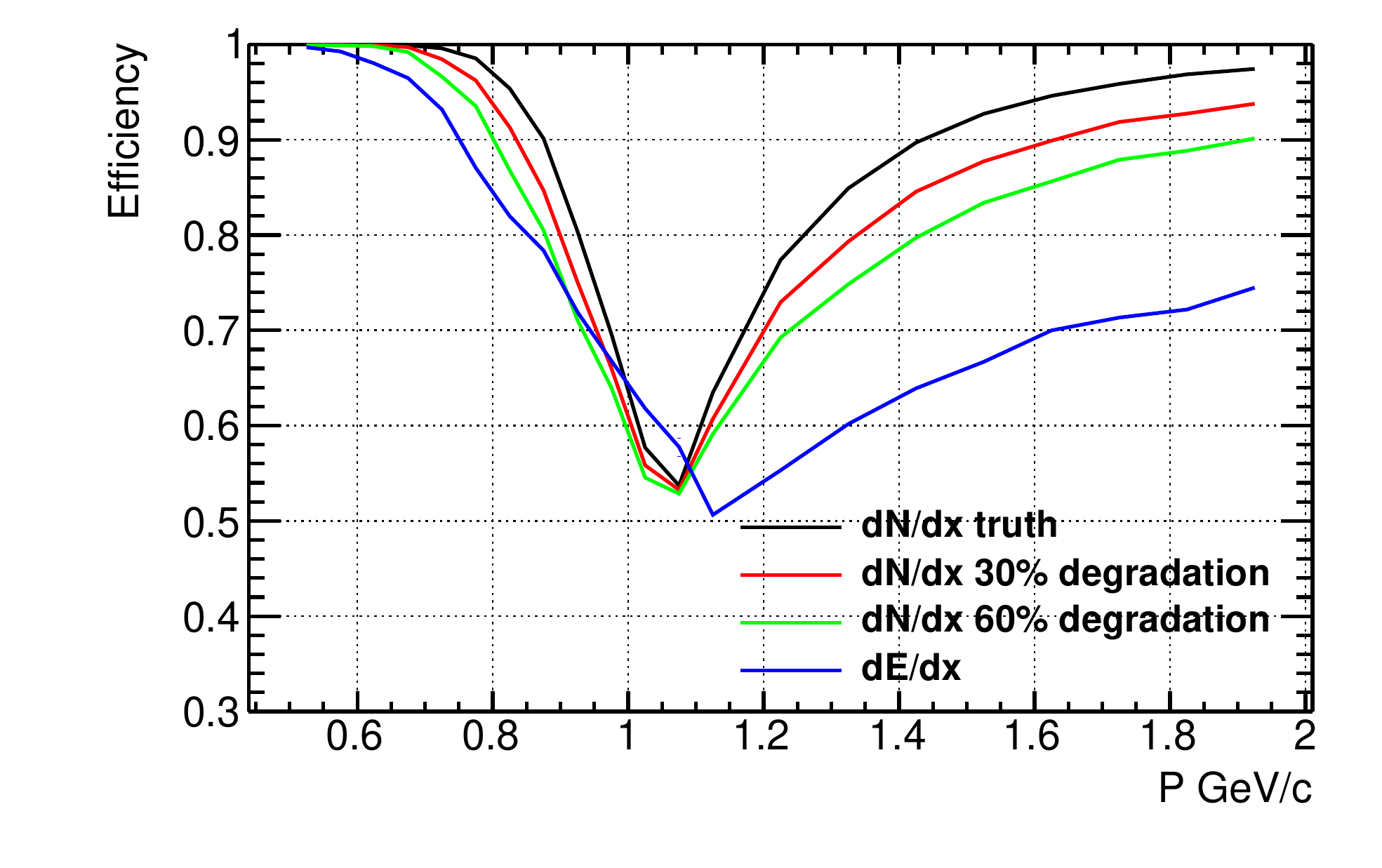}\label{fig:6:3}}
\subfigure[$\pi$ efficiency with TOF]{\includegraphics[width=0.48\linewidth]{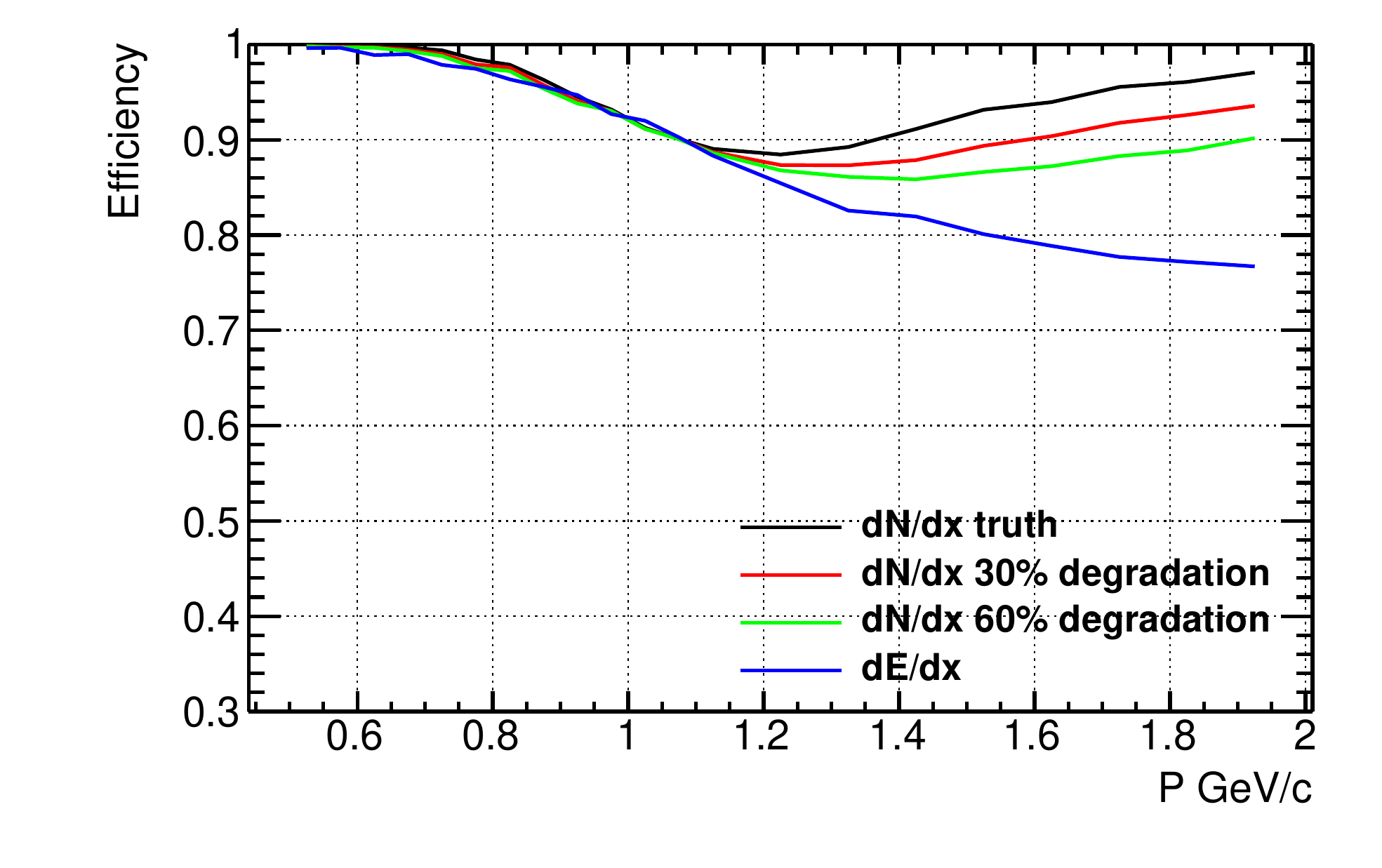}\label{fig:6:4}}
\caption{PID efficiency versus momentum of $\pi$ and K. $dE/dx$ and $dN/dx$ are shown for each sub-plots.} \label{fig:6}
\end{figure*}

The PID efficiencies of $K$ and $\pi$ are displayed in Figure~\ref{fig:6}. Comparing the results with and without TOF, the TOF helps ameliorate PID efficiency from 50\% to around 90\% at momentum around 1.1 GeV/c, but contributes little in high momentum range. 
The combined $dN/dx$ and TOF information provides the best PID efficiency for $K$ and $\pi$, denoted by the line $dN/dx$ in Figure~\ref{fig:6:2},~\ref{fig:6:4}. Compared to the efficiency of $dN/dx$ and $dE/dx$ for $K$ without TOF in Figure~\ref{fig:6:1},
the one with cluster counting of 30\% degradation is 20$\%$ superior to efficiency of energy loss method. With 60\% degradation of $dN/dx$, a gain over 10\% is still possible though in a conservative estimation which considers a peak finding efficiency in noisy background.

\begin{figure}[h]
    \centering
    \includegraphics[width=0.90\linewidth]{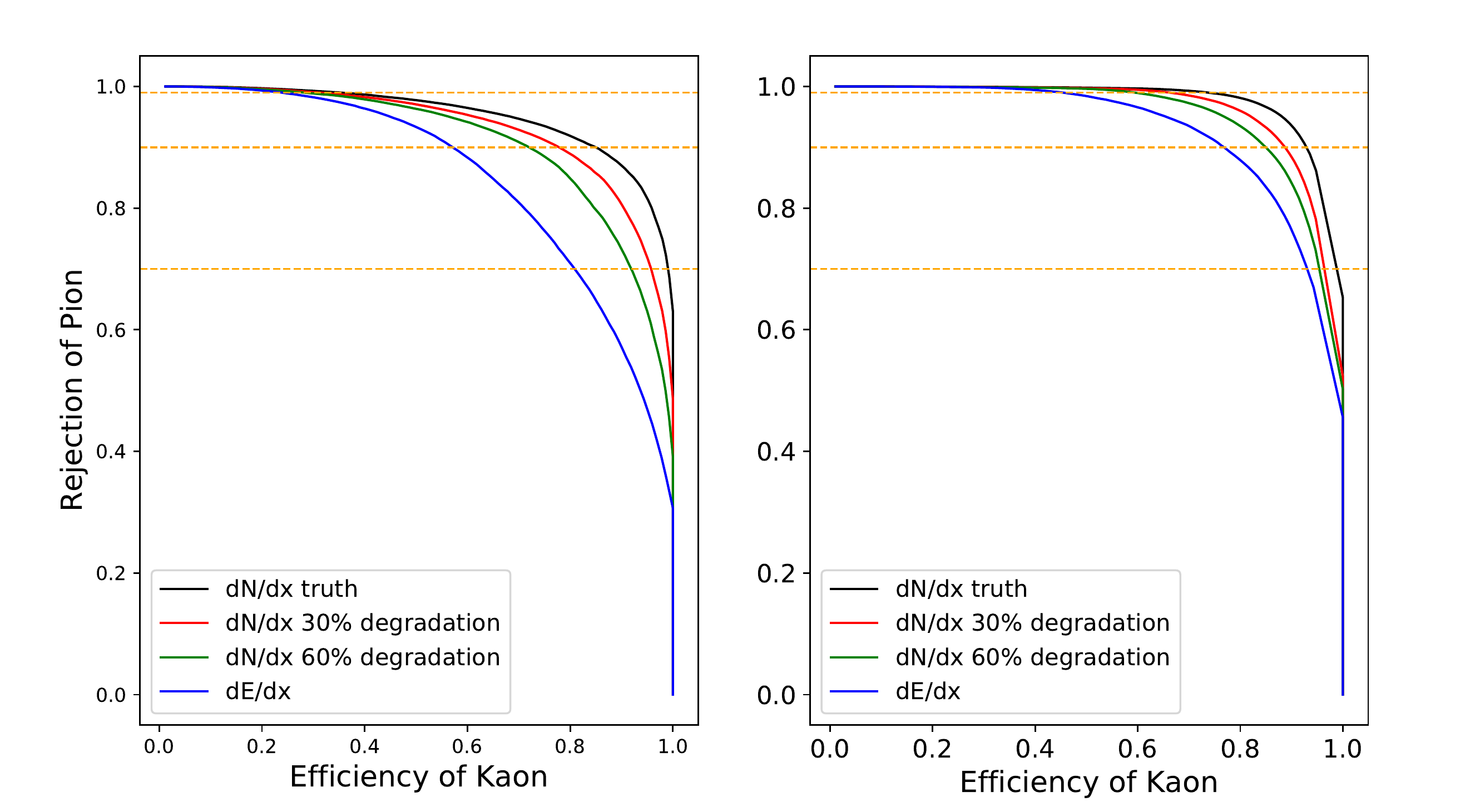}
    \caption{ROC curves made by $K$ selection efficiency versus $K$ selection efficiency. Left: $dN/dx$ and $dE/dx$ without TOF; Right: $dN/dx$ and $dE/dx$ with TOF. Orange dash line indicates $\pi$ rejection efficiency at 70$\%$, 90$\%$ and 99$\%$ respectively.}
    \label{fig:5}
\end{figure}

Another common figure of merit to evaluate the performance of a binary classifier is the receiver operating characteristic (ROC). Larger area under ROC curve indicates a better performance of a classifier. One can also extract information of signal efficiency for a given background rejection level from ROC curve. Figure~\ref{fig:5} plots the $K$ selection efficiency versus $\pi$ rejection efficiency obtained by cutting on $\pi$ probability, for $dE/dx$ and $dN/dx$. The possibilities without TOF and combined with TOF are displayed on the left and right, respectively. The ROC curves of PID methods reflects that combined $dN/dx$ and TOF information provides the best PID performance. We provide benchmarks with $\pi$ rejection efficiency at 70$\%$, 90$\%$ and 99$\%$, shown as orange dash lines. Consequently, with 90\% of $\pi$ rejection in without TOF scenario, an over 80$\%$ of $K$ selection efficiency is expected in $dN/dx$ with theoretical case, increased by at least 20\% with respect to $dE/dx$ .

\clearpage
\section{Conclusion}
This proof-of-concept study demonstrates the profit of cluster counting technique for the \BESthree\, Helium-based drift chamber. 
Specifically, we utilize Garfield++ to generate waveform of the drift chamber signal, different assumption on $dN/dx$ resolution are applied to evaluate PID performance. The separation power between $\pi$ and $K$ with cluster counting foresees a promising improvement over traditional energy loss measurement in the range of momentum above 1.2 GeV/c, gain of a factor of 2.5 and 1.7 are expected in the theoretical case and in the conservative case by 60\% degradation of $dN/dx$ resolution, respectively. 
Furthermore, $dN/dx$ model is parameterized in BOSS analysis toolkit. The PID efficiency combining the drift chamber and the time-of-flight counters is evaluated using Monte Carlo simulation. The results show that the PID efficiency of $\pi$ and $K$ is enhanced significantly. To achieve an excellent PID capability delivered by cluster counting technique, high performance of front-end readout electronics with low noise is necessary to resolve signal pluses from different primary clusters. Aside from the simulation study, prototype tests are expected to validate the feasibility of cluster counting technique.

\label{sec:4}

\acknowledgments

This research was funded by National Key R\&D Program of China under Contracts Nos. 2020YFA0406304, National Natural Science Foundation of China (NSFC) under Contract Nos. U1832204, 12275296, 11521505 and National 1000 Talents Program of China.


\begin{thebibliography}{99}

    
\bibitem{20097}
    BESIII Collaboration,
    \emph{The construction of the BESIII experiment},
    \href{https://doi.org/10.1016/j.nima.2008.08.072}{\emph{Nucl. Instrum. Meth. A} {\bf 598} (2009) 7-11}
    
\bibitem{BESIII:2020nme}
    Ablikim, M. et al.,
    \emph{Future Physics Programme of BESIII}
    \href{http://dx.doi.org/10.1088/1674-1137/44/4/040001}{\emph{Chin. Phys. C} {\bf 44} (2020) 040001}
    

    
\bibitem{Shengsen_2012}
	Sun Shengen,
	\emph{Time calibration for barrel {TOF} system of {BESIII}}
	\href{https://doi.org/10.1088/1742-6596/396/2/022051}{\emph{Journal of Physics} {\bf 396} (2012) 022051}

\bibitem{Guo2017}
   Guo, Ying-Xiao et al.,
   \emph{The study of time calibration for upgraded end cap TOF of BESIII}
   \href{https://doi.org/10.1007/s41605-017-0012-4}{\emph{Radiation Detection Technology and Methods } {\bf 1} (2017) 15}

\bibitem{4329616}
    Walenta, A. H.,  
    \emph{The Time Expansion Chamber and Single Ionization Cluster Measurement}
    \href{https://doi.org/10.1109/TNS.1979.4329616}{\emph{IEEE Transactions on Nuclear Science} {\bf 26} (1979) 73-80}    

\bibitem{LAPIQUE1980297}
    F. Lapique and F. Piuz,
    \emph{Simulation of the measurement by primary cluster counting of the energy lost by a relativistic ionizing particle in argon}
    \href{https://doi.org/10.1016/0029-554X(80)90744-2}{\emph{Nuclear Instruments and Methods} {\bf 175} (1980) 297-318}
    

\bibitem{Abada2019}
   Abada, A, et al.,
   \emph{FCC-ee: The Lepton Collider},
   \href{https://doi.org/10.1140/epjst/e2019-900045-4}{\emph{The European Physical Journal Special Topics} {\bf 228} (2019) 261-623}

\bibitem{cepc}
   M. Ahmad et al.,
   \emph{CEPC-SPPC Pre-CDR.},
   \href{}{Chap. 6, (2018)}

\bibitem{Garfield}
    Garfield++,
    \href{https://garfieldpp.web.cern.ch/garfieldpp}{https://garfieldpp.web.cern.ch/garfieldpp}

\bibitem{boss}
    DENG Zi-Yan et al.,
    \emph{Object-Oriented \BESthree\, detector simulation system},
    \href{http://cpc.ihep.ac.cn/article/id/283d17c0-e8fa-4ad7-bfe3-92095466def1}{\emph{Chinese Physics C} {\bf 30} (2006) 371-377}

\bibitem{root}
    Rene Brun and Fons Rademakers, 
    \emph{ROOT - An Object Oriented Data Analysis Framework},
    \href{}{\emph{Proceedings AIHENP'96 Workshop, Lausanne, Sep. 1996, Nucl. Inst. \& Meth. in Phys. Res. A} {\bf 389} (1997) 81-86}

\bibitem{Pfeiffer:2018yam}
    Pfeiffer, Dorothea et al.,
    \emph{Interfacing Geant4, Garfield++ and Degrad for the Simulation of Gaseous Detectors},
    \href{https://doi.org/10.1016/j.nima.2019.04.110}{\emph{Nucl. Instrum. Meth. A} {\bf 935} (2019) 121-134}
    
\bibitem{AGOSTINELLI2003250}
    S. Agostinelli, et al.,
    \emph{Geant4—a simulation toolkit}
    \href{https://doi.org/10.1016/S0168-9002(03)01368-8}{\emph{Nucl. Instrum. Meth. A} {\bf 506} (2003) 250-303}

\end{thebibliography}

\end{document}